\documentclass[twocolumn, superscriptaddress]{revtex4-1}

\usepackage[paperwidth=210mm,paperheight=297mm,centering,hmargin=1.5cm,vmargin=2.5cm]{geometry}

\usepackage[utf8]{inputenc}
\usepackage[english]{babel}
\usepackage[T1]{fontenc}

\usepackage{graphicx}
\usepackage{float}
\usepackage{xcolor}

\usepackage{amsmath,amsfonts,amssymb}

\usepackage{ntheorem}
  \theoremseparator{.}
  \newtheorem{hyp}{Hypothesis}

\usepackage{hyperref}

\begin{document}
\title{Impossibility of memory in hidden-signaling
models for quantum correlations}
 
\author{Ignacio Perito}
\affiliation{Departamento de Física, Facultad de Ciencias Exactas y Naturales, Universidad de Buenos Aires, 1428 Buenos Aires, Argentina}
\affiliation{CONICET-UBA, Instituto de Física de Buenos Aires (IFIBA), 1428 Buenos Aires, Argentina}
\author{Guido Bellomo}
\affiliation{Departamento de Física, Facultad de Ciencias Exactas y Naturales, Universidad de Buenos Aires, 1428 Buenos Aires, Argentina}
\affiliation{CONICET-UBA, Instituto de Investigación en Ciencias de la Computación (ICC), 1428 Buenos Aires, Argentina}
\author{Daniel Galicer}
\affiliation{Departamento de Matemática, Facultad de Ciencias Exactas y Naturales, Universidad de Buenos Aires, 1428 Buenos Aires, Argentina, and IMAS-CONICET}
\author{Santiago Figueira}
\affiliation{CONICET-UBA, Instituto de Investigación en Ciencias de la Computación (ICC), 1428 Buenos Aires, Argentina}
\affiliation{Departamento de Computación, Facultad de Ciencias Exactas y Naturales, Universidad de Buenos Aires, 1428 Buenos Aires, Argentina}
\author{Augusto J. Roncaglia}
\affiliation{Departamento de Física, Facultad de Ciencias Exactas y Naturales, Universidad de Buenos Aires, 1428 Buenos Aires, Argentina}
\affiliation{CONICET-UBA, Instituto de Física de Buenos Aires (IFIBA), 1428 Buenos Aires, Argentina}
\author{Ariel Bendersky}
\affiliation{CONICET-UBA, Instituto de Investigación en Ciencias de la Computación (ICC), 1428 Buenos Aires, Argentina}
\affiliation{Departamento de Computación, Facultad de Ciencias Exactas y Naturales, Universidad de Buenos Aires, 1428 Buenos Aires, Argentina}

\date{\today}

%%% ABSTRACT
\begin{abstract}
We consider a toy model for non-local quantum correlations in which nature resorts to some form of hidden signaling (i.e., signaling between boxes but not available to the users) to generate correlations. We show that if such a model also had memory, the parties would be able to exploit the hidden-signaling and use it to send a message, achieving faster-than-light communication. Given that memory is a resource easily available for any physical system, our results add evidence against hidden signaling as the mechanism behind nature's non-local behavior. 
\end{abstract}  

\maketitle

%%% INTRODUCTION
\section{Introduction}

Since the seminal work by Bell~\cite{Bell1964}, it is known that some correlations allowed by quantum mechanics cannot be obtained by means of local hidden-variable models. In other words, quantum mechanics exhibits correlations that are impossible to describe within a model that is both local and realistic. Another feature of quantum correlations is that of being non-signaling, i.e.,  they cannot give place to faster-than-light communication. 
Thus, quantum correlations cannot be explained via  a classical model without communication.
In fact, several studies quantify the amount of classical communication that is needed to reproduce Bell correlations   \cite{Toner2005,regev2010simulating,shi2008tensor,degorre2009communication}.
Here we are interested in possible mechanisms by which nature gives rise to these kinds of non-local correlations. A possible approach to account for a deterministic description of non-local correlations is to resort to some signaling mechanism. However, since quantum correlations are non-signaling, this mechanism should  be restricted to the hidden variables level, not reaching the phenomenological one. 
In this respect, a paradigmatic example of a deterministic non-local theory that exhibits signaling at the hidden-variable level is Bohmian mechanics~\cite{Bohm1952}.

In a standard Bell scenario, a source prepares a pair of particles in some state, and then each particle is sent to two distant observers which implement random measurements chosen from a finite set. The experiment is repeated many times and the observers collect the data. This situation is usually represented by two abstract boxes that receive inputs and give some outputs as a result. In Ref.~\cite{Bendersky2017},  a deterministic model with hidden-signaling between these boxes was considered and it was shown that if the outputs of the boxes were generated using a computable function, the parties could signal each other. Thus, deterministic models reproducing non-local correlations must be uncomputable. One could also consider situations in which the outputs of the boxes are affected by the results of previous rounds, that is, a situation where the outputs are conditioned by the memory of the devices. While not in the context of a deterministic description of non-local correlations,  Bell scenarios with memory were also extensively studied \cite{Barrett2002, PhysRevLett.116.230402, PhysRevLett.110.010503,pironio2010random,pironio2013security}. In Ref.~\cite{Barrett2002}, it
was shown that even in the presence of memory between rounds a sufficiently large violation of a Bell inequality suffices to prove non-locality (see also Refs.~\cite{Gill2003}).

In this paper we consider a model for non-local correlations that combines  both situations: hidden-signaling (i.e., a hidden communication among the devices that is not available to the agents) and memory. We demonstrate that the presence of memory  turn hidden-singnaling into a resource to instantly communicate information, since it allows the agents to signal each other. Given that memory is a common resource in nature, these results add evidence against hidden-signaling as a model for non-local correlations. Notably, this scenario differs from the one considered in Ref.~\cite{Bendersky2017}, since we allow computable and uncomputable boxes. In this case, we show that even if the boxes produce their outputs in an uncomputable fashion, hidden-signaling is not allowed unless there is some strange \textit{self-censorship mechanism} by which nature forbids itself from keeping records of the past. More specifically, our result shows that if nature uses any kind of hidden signaling between the parties, then it has to be unable to remember the previous rounds (otherwise, this hidden-signaling could be extracted by the parties to send information superluminally.)

The paper is organized as follows: In Sec.~\ref{sec:scenario}, we introduce the model and the hidden signaling mechanism. In Sec.~\ref{sec:main}, first we check that it is possible to reproduce non-local 
correlations with hidden-signaling schemes, and then we report our main results showing that hidden-signaling plus memory allow the parties to signal each other. In Sec.~\ref{sec:protocolo}, we describe a sampling protocol needed to achieve the signaling. Finally, we give a summary of our results.

%%% SCENARIO
\section{Scenario}
\label{sec:scenario}

We consider a Bell-like scenario with two parties, Alice and Bob, each with access to a box with two measurement choices and two possible outcomes. We define $x_n$ and $y_n$ as the inputs (measurement choices) for Alice and Bob, respectively, in the $n$-th round. Similarly, $a_n$ and $b_n$ are the outputs (measurement results) obtained by Alice and Bob in that round. Before moving on, we should be clear about our notation. As we will be dealing with sequences of inputs and outputs (in the sense that we will need to refer to the inputs/outputs of previous or forthcoming rounds), we will refer to $p(a_n,b_n|x_n,y_n)$ to the probability of observing the outputs $a_n, b_n$ on the $n$--th round of an experiment given inputs $x_n,y_n$ on the same round. We can also write probabilities such as $p(a_n,b_{n-1}|x_n,y_{n-1})$ as the probability of observing the symbol $a_n$ as Alice's output on the $n$--th round and $b_{n-1}$ as Bob's output on the previous round, while having as input for Alice $x_n$ on the $n$--th round and for Bob $y_{n-1}$ in the previous round. The symbols $a$ and $b$ will be used for outputs (for Alice and Bob respectively), with subindexes for the round number, and $x$ and $y$ for the inputs, with corresponding subindexes.

We make the usual assumption that the relative frequencies of the outputs, given the inputs, are independent of the round. That is, we have
\begin{equation}
  p( a_n,b_n | x_n,y_n ) =  p(a_{m},b_{m}|x_{m},y_{m}) \quad \forall n,m \in \mathbb{N} \,,
\end{equation}
for some fixed quantum and non-local distribution $p(a,b|x,y)$. In particular, we are concerned with possible mechanisms that nature could use to produce non-local, but also non-signaling, distributions. A distribution is non-signaling when its marginals are well defined, that is
\begin{equation} \label{eq:ns}
  p( a_n| x_n ) = \sum_{b}   p( a_n,b_n | x_n,y_n ) 
\end{equation}
is independent of $y$, and similarly for Bob's side \cite{Popescu1994}.

It is known that deterministic models that reproduce these kind of correlations require the existence of \textit{hidden signaling} between the parties \cite{maudlin1992d, cleve1997substituting,steiner2000towards, csirik2002cost, bacon2003bell, toner2003communication, Bendersky2017}. However, it is important to remark that the existence of a hidden-signaling mechanism does not trigger, in principle, any conflict with special relativity. Whether this assumption can have undesirable effects at the observational level (i.e, whether there is effective \textit{faster-than-light} information transfer) depends upon the nature of the deterministic functions that describe the inner working of the boxes. Undesirable effects arise, for example, when the outputs are computable functions of both inputs and of the round number, as we have already mentioned~\cite{Bendersky2017}. Now we will analyze when we allow nature to keep record of the hidden-signaling of previous rounds and use this information to produce the next output.

\begin{figure}[t!]
    \centering
    \includegraphics[width=1\linewidth]{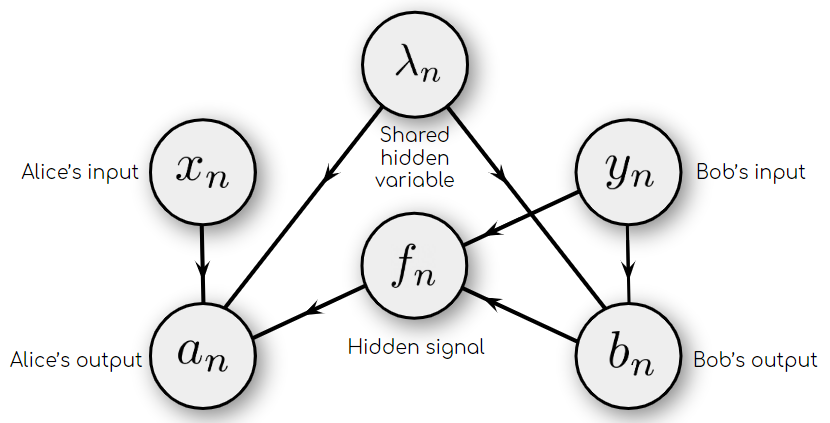}
    \caption{Directed acyclic graph depicting the causal structure of the deterministic model. Alice and Bob run a Bell-like experiment by implementing measurements on their distant labs. On each round, a hidden signal that depends on Bob's input and/or output is available to Alice's box. Both boxes also share the value of a hidden variable $\lambda_n$. }
    \label{fig:scheme}
\end{figure}

Figure \ref{fig:scheme} shows the local and deterministic model that we consider in this paper using the language of causal networks \cite{pearl2009causality, chaves2015information, henson2014theory, PhysRevLett.114.140403, wood2015lesson}, that is, directed acyclic graphs indicating causal relations between variables: an arrow from node $A$ to node $B$ means that variable $B$ is causally  linked to variable $A$. Alice's box generates its output using the value of its input, a hidden variable shared with Bob, and the value of a function $f: \{ 0,1 \} ^2 \rightarrow \{ 0,1 \}^2$ of the two local variables on Bob's side that is hiddenly signaled. We will adopt the natural assumption of uniformity, that is, the information being signaled between the boxes (not known \textit{a priori}) is the same for all behaviors $p(ab|xy) \in \mathcal{Q}$, the convex set of quantum behaviors. 
For simplicity, we will first consider the simplest case, when there is only one bit of one-way hidden signaling between the boxes (that is enough to reproduce non locality in the memoryless case) and nature uses only the preceding round to produce the next pair of outputs. The generalization to signaling of more than one bit and/or memory of multiple rounds will be straightforward. 

\begin{figure}[b!]
    \centering
    \includegraphics[width=0.85\linewidth]{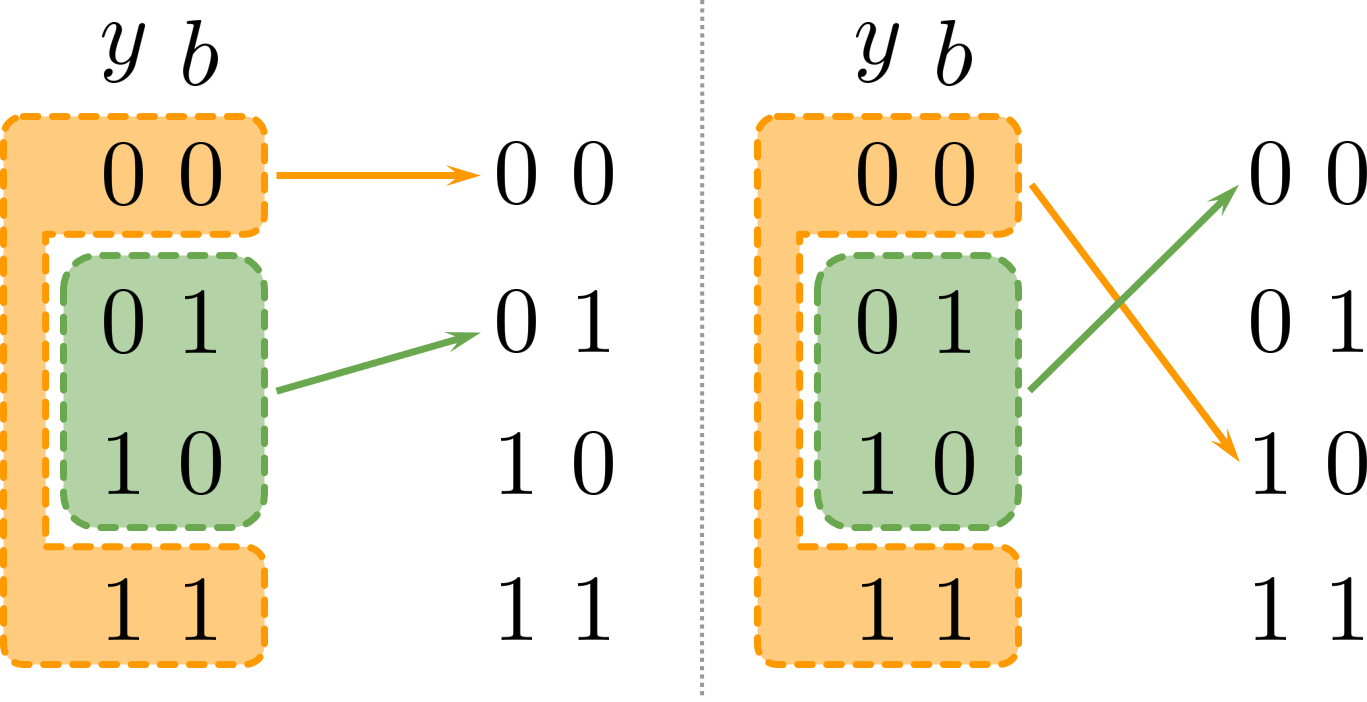}
    \caption{Two physically equivalent functions $f: \{ 0,1 \} ^2 \rightarrow \{ 0,1 \}^2$. Left and right relations are different, but they both correspond to the case in which the information being transmitted from Bob's side is $f(y,b) = y \oplus b$.
    %, as well as all the other ten possible choices for the ending points of orange and green arrows. 
 The function is determined once its domain is partitioned, therefore the corresponding values on the target set have no physical meaning.}
    \label{fig:partitions}
\end{figure}

Notice that all $256$ possible functions $f: \{ 0,1 \} ^2 \rightarrow \{ 0,1 \}^2$ can be split in classes that correspond to physically equivalent situations. Each of these classes can be thought of as a partition of the set $\{ 0,1 \} ^2$. Figure \ref{fig:partitions} schematizes this fact for the case $f(y,b) = y \oplus b$, where $\oplus$ denotes modulo $2$ addition. There are $15$ possible partitions of a set of four elements, so there are $15$ physically relevant functions that can be signaled between the boxes, and one of them corresponds to the constant function, which does not carry any information, so we are left with $14$ relevant functions.

\begin{figure}[t!]
    \centering
    \includegraphics[width=0.8\linewidth]{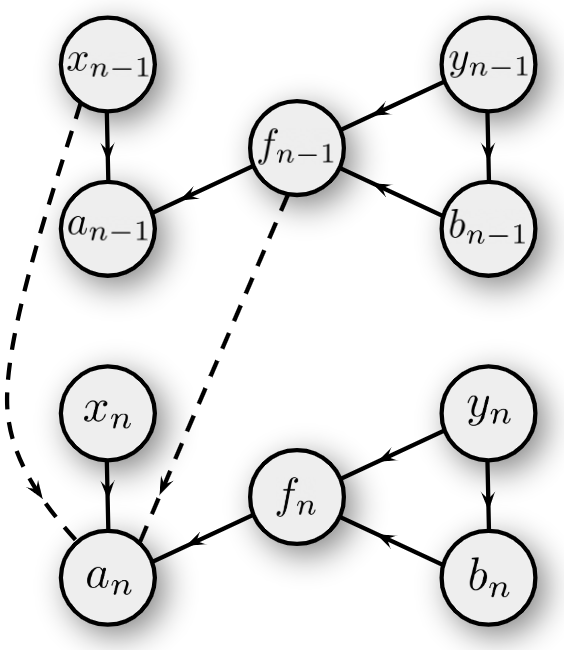}
    \caption{Directed acyclic graph depicting the causal structure of the hidden signaling scenario when memory effects are present. The variables that enter Alice's device in the round $n-1$ have a causal  effect on Alice's output in the $n$-th round (memory effects are represented by dashed lines). Given that we have traced out hidden variables, edges between nodes should be thought as probabilistic (rather than deterministic) causal links. }
    \label{fig:memorynetwork}
\end{figure}

Suppose that we look at the local statistics on Alice's side. The fact that there is hidden signaling between the boxes in a given round implies that Alice's box has access to some kind of information about Bob's input and/or output on that round. If nature keeps record of the hidden signaling of the previous round and uses it to generate the next output, then the following hypothesis must hold:
\begin{hyp} [Memory]\label{hyp:first}
  The conditional probability distribution $p\left(a_n | x_n , x_{n-1}, f(y_n,b_n ), f(y_{n-1},b_{n-1} ) \right)$ has a non-trivial dependence on $f(y_{n-1},b_{n-1} )$.
\end{hyp}
For $p\left(a_n | x_n , x_{n-1}, f(y_n,b_n ), f(y_{n-1},b_{n-1} ) \right)$ to have a non-trivial dependence on $f(y_{n-1},b_{n-1} )$ is to say that there exist $a_n$, $x_n$, $x_{n-1}$, $y_n$ and $b_n$ such that $p\left(a_n | x_n , x_{n-1}, f(y_n,b_n ), f(y_{n-1},b_{n-1} ) \right) \neq p\left(a_n | x_n , x_{n-1}, f(y_n,b_n ), f(y'_{n-1},b'_{n-1} ) \right)$, at least for a pair $y'_{n-1} , b'_{n-1}$ such that $f(y'_{n-1},b'_{n-1} ) \neq f(y_{n-1},b_{n-1} )$. Figure \ref{fig:memorynetwork} shows the causal structure of the hidden signaling scenario when memory effects, in the sense of Hypothesis \ref{hyp:first}, are present.

If nature uses a hidden signaling mechanism to produce the non-local correlations then, given that memory is a widely available resource in nature, it must also be compatible with special relativity when 
the boxes are allowed to keep record of this signaling. However, we are going to show that Hypothesis~\ref{hyp:first}, when specializing on some given non-local quantum distributions, leads to a scenario in which Alice and Bob could communicate superluminally.

Now, we proceed to study what happens when Hypothesis~\ref{hyp:first} holds for particular instances of the function $f$. 
We will start by considering one bit functions corresponding to either Bob's input or output. 
Those cases cover the key points of our argument and, later, can easily be extended to more general functions.

%%% SCENARIO
\section{One bit of hidden signaling from Bob's side}
\label{sec:main}

Our argument has two main ingredients: first we check that hidden signaling of a given function is enough to reproduce non-local correlations in a deterministic memoryless scenario, and then we show that if memory is allowed, this resource would allow Alice and Bob to communicate instantly, thus reaching a contradiction with special relativity. In the following two sections we follow this strategy in detail for the cases in which the function $f$ that is signaled
corresponds to either Bob's input or output. Then we  generalize the idea for broader functions.

%%%
\subsection{Signaling of Bob's input}\label{sec:input}

Let us first consider the case in which Alice's box receives in each round the bit associated to the value of Bob's input. As we said, we will first show that hidden signaling of Bob's input is enough to obtain any non-local behavior in a deterministic fashion (that is, there is a hidden variable model that reproduces non-locality if Alice's box has access to Bob's input on each round). Then, we will proceed to study the implications of the existence of memory in such an scenario, that is, we will assume that Hypothesis~\ref{hyp:first} holds with $f(y,b) = y$.
This example covers all the functions $f: \{ 0,1 \} ^2 \rightarrow \{ 0,1 \}^2$ for which the domain is split into two subsets, $y=0$ and $y=1$, with two elements each. Our goal is to show that  Alice, by looking at her local data in a given round, can infer something about Bob's input of the previous round.

The first step is to show that any non-signaling distribution $p(ab|xy)$ can be reproduced by the scheme of Figure~\ref{fig:scheme} with $f(b,y) = y$. In the $(2,2,2)$ scenario (that is a scenario with two parties, each one with two inputs and two outputs), an arbitrary non-signaling distribution can be written as a convex combination of the $24$ vertices of the non-signaling polytope. Out of those $24$ vertices, $16$ correspond to the deterministic behaviors, which are trivial to reproduce in this scheme, given that they require neither hidden signaling nor a shared hidden variable between the parties.  The remaining vertices are the eight relabelings of inputs and outputs of the well-known PR boxes \cite{Popescu1994}, which satisfy that, locally, the boxes are just unbiased coins (namely, $p(a|x)=p(b|y)=1/2$) and
\begin{equation}
a \oplus b = xy \, ,
\label{eq:PRcond}
\end{equation}
for every round. If we are able to reproduce each of these behaviors with our scheme, then we can reproduce any non-signaling distribution, because all vertices can be implemented in this way and any convex combination can be simulated by conditioning the response of the boxes to the value of a shared hidden variable. 

To reproduce PR boxes under the scheme  of Fig.~\ref{fig:scheme}, consider that $\lambda_n$ is, for all $n$, a random unbiased bit and that Bob's output is the following deterministic function of his input and the hidden variable
\begin{equation}
b_n = y_n \oplus \lambda_n \, ,
\end{equation}
which automatically satisfies $p(b|y)=1/2$. Now, since in each round Alice’s box has access to $y_n$ due to the hidden signaling, its output can be given by the following deterministic function
\begin{equation}
a_n = x_n y_n \oplus y_n \oplus \lambda_n \, .
\end{equation}
At first sight, the fact that $a_n$ depends explicitly upon $y_n$ could seem to violate the non-signaling condition, but it is easy to check that $p(a_n|x_n y_n) = 1/2$, thus the local statistics is unbiased on both sides. Now, from the last two equations, we have
\begin{equation}
a_n \oplus b_n = (y_n \oplus \lambda_n) \oplus (x_n y_n \oplus y_n \oplus \lambda_n) = x_n y_n \, ,
\end{equation} 
and then the PR condition Eq.~\eqref{eq:PRcond} is satisfied for every round. We have shown that the correlations given by PR boxes can be reproduced by a deterministic model with one-way hidden signaling of one side's input. As we said before, the other non-local vertices can be obtained from this model by re-labeling inputs and outputs; and the local vertices are trivially reproduced in this scheme, so we conclude that any non-signaling correlation can be obtained in this manner. In the next step, we show that the presence of memory  allow the parties to extract the hidden-signaling and use it for communication.

Hypothesis \ref{hyp:first} in the current situation reduces to the statement that
\begin{equation}
\begin{split}
& p(a_n | x_n = \tilde{x} , x_{n-1} = \tilde{\tilde{x}} , y_n = \tilde{y} , y_{n-1} = y)  \\
&\neq  p(a_n |  x_n = \tilde{x} , x_{n-1} = \tilde{\tilde{x}} , y_n = \tilde{y} , y_{n-1} = y') 
\end{split}
\label{memoryinput}
\end{equation}
if $y \neq y'$, for some fixed $\tilde{x}$, $\tilde{\tilde{x}}$, and $\tilde{y}$. Now, if Alice and Bob have knowledge of the probability distribution $p(a_n | x_n , x_{n-1} , y_n , y_{n-1})$ (something that is possible, as we will see in Sec.~\ref{sec:protocolo}), then they can agree to do the following: On odd rounds, Alice chooses $\tilde{x}$ as her input and Bob chooses $\tilde{y}$. On even rounds, Alice chooses $\tilde{\tilde{x}}$ and Bob chooses the bit he wants to send to Alice. In this way, the probability distribution of Alice's outputs in odd rounds is given by
\begin{equation}
\begin{split}
p ( a_{2n+1} = 0 | \tilde{x} , \tilde{\tilde{x}} , \tilde{y} , y_{2n} = 0 ) & \equiv \alpha \, , \\
p ( a_{2n+1} = 0 | \tilde{x} , \tilde{\tilde{x}} , \tilde{y} , y_{2n} = 1 ) & \equiv \beta \, ,
\end{split}
\end{equation}
where Eq.~\eqref{memoryinput} ensures that $\alpha\neq\beta$ are two probabilities known by Alice and Bob. If they stick to this strategy, in the limit of infinitely many rounds Alice will see that the fraction of times that her device outputs a zero in the odd rounds is either $\alpha$ or $\beta$; and from this information she can tell which bit Bob is trying to send. As we will see in Sec.~\ref{sec:protocolo}, if they are restricted to a finite number of rounds, Alice can use this information to guess Bob's input in even rounds with some probability of success greater than $1/2$, and hence they are able to communicate within this scenario. Of course,  this conflicts with special relativity, because one can always fix the distance between the parties and the time between rounds in such a way that this communication is faster-than-light.  

%%%
\subsection{Signaling of Bob's output}\label{sec:output}

Now we consider the situation in which Hypothesis~\ref{hyp:first} holds with $f(y,b) = b$. That is, Alice's box receives Bob's output. We will start by showing that the hidden signaling of this variable gives rise to a plausible deterministic model of non-local correlations. Like in the previous case, it is sufficient to show that we can reproduce PR boxes within this scheme. Again, we can take $\lambda_n$ to be a fair coin tossing for all $n$ and set
\begin{equation}
b_n = y_n \oplus \lambda_n \; \forall n \in \mathbb{N} \, ,
\label{eq:bnoutputcase}
\end{equation}
which satisfies that $b_n$ is locally a random unbiased bit independent of $y_n$. The PR condition Eq.~\eqref{eq:PRcond} requires that
\begin{equation}
a_n = x_n y_n \oplus b_n \; \forall n \in \mathbb{N} \ . 
\end{equation}
Given that Alice's box has access to $\lambda_n$ and $b_n$, and Eq.~\eqref{eq:bnoutputcase} allows us to write $y_n$ as $b_n \oplus \lambda_n$, this can be achieved by setting
\begin{equation}
a_n = x_n ( b_n \oplus \lambda_n ) \oplus b_n \; \forall n \in \mathbb{N} \,  ,
\end{equation}
which automatically matches the PR condition for every round and also satisfies the condition of locally unbiased statistics for Alice's variables.

In this case, Hypothesis \ref{hyp:first} says that
\begin{equation}
\begin{split}
& p(a_n | x_n = \tilde{x} , x_{n-1} = \tilde{\tilde{x}} , b_n = \tilde{b} , b_{n-1}=b)  \\
&\neq  p(a_n |  x_n = \tilde{x} , x_{n-1} = \tilde{\tilde{x}} , b_n = \tilde{b} , b'_{n-1}=b') 
\end{split}
\label{memoryoutput}
\end{equation}
if $b \neq b'$, for some fixed $\tilde{x}$, $\tilde{\tilde{x}}$ and $\tilde{b}$. Unlike the previous case, where memory immediately  allowed Alice to infer something about Bob's input, here we have that the bias in the local distribution of Alice is modified by Bob's output, which is something that he, in principle, has no control over. 
This makes our argument a little more convoluted; and we will need to fix the distribution $p(ab|xy)$ in a convenient manner in order to reach a communication protocol.

First, note that there are distributions for which Bob has perfect control over his outputs: the vertices of the local polytope are deterministic behaviors, which include, for example, the case of boxes that  output just a copy of the input. Furthermore, in the $(2,2,2)$ scenario, we know that the local polytope has $24$ facets, eight of which correspond to CHSH-like inequalities \cite{Froissart1981,Brunner2014} and that for any local behavior lying on such a facet, there exist non-local and quantum behaviors arbitrarily close to it.

 Let $p_v (ab | xy)$ be a vertex of the local polytope such that $p_v ( b | y) = \delta_{by}$ (that is, Bob's output is just a copy of his input). It is easy to check that this behavior lies on one of the eight CHSH facets. We then consider a scenario where the round-to-round distribution is such that
\begin{equation}
p(a_n b_n | x_n y_n ) \simeq p_v (a_n b_n | x_n y_n ) \; \forall n \in \mathbb{N} \, ,
\end{equation} 
where the $\simeq$ symbol means that $p(a_n b_n | x_n y_n )$ is a quantum, non-local distribution arbitrarily close to $p_v (a_n b_n | x_n y_n )$. In this case, we can write
\begin{equation}
\begin{split}
p ( a_{2n+1} = 0 | \tilde{x} , \tilde{\tilde{x}} , \tilde{b} , b_{2n} = 0 ) & \equiv \alpha \, , \\
p ( a_{2n+1} = 1 | \tilde{x} , \tilde{\tilde{x}} , \tilde{b} , b_{2n} = 1 ) & \equiv \beta \, ,
\end{split}
\end{equation}
where, again, we have $\alpha \neq \beta$ by hypothesis. If, like in the previous case, Alice always chooses $\tilde{x}$ as her input in odd rounds and $\tilde{\tilde{x}}$ in the even ones, we can write the previous lines in their frequentist versions
\begin{equation}
\begin{split}
\lim_{n\rightarrow\infty} & \frac{\# \left\{ i<n \, / \, (a_{2i+1}, b_{2i+1} , b_{2i} ) = ( 0,\tilde{b},\tilde{\tilde{b}})  \right\} }{\#\left\{ i<n \, / \,  (b_{2i+1} , b_{2i}) = ( \tilde{b},\tilde{\tilde{b}} ) \right\} } = \alpha \, , \\ 
\lim_{n\rightarrow\infty} & \frac{\# \left\{ i<n \, / \, (a_{2i+1}, b_{2i+1} , b_{2i} ) = ( 1,\tilde{b},\tilde{\tilde{b}})  \right\} }{\#\left\{ i<n \, / \,  (b_{2i+1} , b_{2i}) = ( \tilde{b},\tilde{\tilde{b}} ) \right\} } = \beta \, ,
\end{split}
\end{equation}
where $\# \star $ is the cardinal number of the set $\star$. Because of the particular distribution we are considering, the rounds on which Bob's box outputs $0$ or $1$ are (almost) the same rounds in which those are, respectively, his inputs. Then, we can write
\begin{equation}
\begin{split}
p ( a_{2n+1} = 0 | \tilde{x} , \tilde{\tilde{x}} , y_{2n+1} = \tilde{b} , y_{2n} = 0 ) & \simeq \alpha \, , \\
p ( a_{2n+1} = 1 | \tilde{x} , \tilde{\tilde{x}} , y_{2n+1} = \tilde{b} , y_{2n} = 1 ) & \simeq \beta \, ,
\end{split}
\end{equation}
and, as before, the $\simeq$ symbol means that both members in the equation can be made arbitrarily close by choosing the round to round distribution properly. At this point, we see that if Alice and Bob are able to reconstruct those probability distributions (which, as we will show below, they are), then Alice can proceed in the same way as in the previous subsection and the same contradiction with special relativity arises.  

\subsection{Signaling of other one bit functions}
\label{sec:other}

From the two previous sections, it is clear that if we want to show that a  deterministic model for non-local correlations based on hidden signaling is not compatible with the existence of memory and the restrictions imposed by causality, there are two conditions that the function being signaled must satisfy. First, a memoryless hidden variable model of non-local correlations using this hidden signaling must exist (otherwise, there is no interest in studying its physical implications when adding memory to it). Second, it seems necessary to find certain non-local and quantum distributions $p(ab|xy)$ such that Bob has control over the value of the function being signaled, as we did in the case in which $f(b,y) = b$. This is crucial to make communication possible between the parties, but it was not required for $f(b,y) = y$, rendering the result independent of the particular distribution. In that case, the bit that is signaled is directly chosen by Bob (this is the main difference between those two cases). 

The one-bit functions that were not considered so far are the XOR function $f(y,b) = y \oplus b$ (that is, one bit indicating whether Bob's input and output are equal or not); and all possible relabelings of the AND function, $f(y,b) = yb$. For the XOR function, it is easy to check that if Alice and Bob share the result of a fair coin $\lambda$ and  set
\begin{equation}
\begin{split}
b &= \lambda \, , \\
a &= x \left[ \lambda \oplus (y\oplus b )\right] \oplus \lambda \, ,
\end{split}
\end{equation}
on each round, then the round to round distribution $p(ab|xy)$ is the same as the one given by PR boxes. Therfore, a shared hidden variable and the signalization of $y\oplus b$ from Bob's side to Alice's side is enough to reproduce any non-local behavior (by the same arguments of the previous sections). In addition, if we consider a local vertex for which Bob's box always outputs the same value independently of the input, then Bob can clearly decide the value of $y \oplus b$ on each round. Again, this vertex lies on a CHSH facet, so taking a quantum non-local distribution arbitrarily close to this vertex, we reach a contradiction with causality from Hypothesis \ref{hyp:first}, and we can proceed exactly in the same way as in Sec. \ref{sec:output}. 

For the AND function ($f(y,b) = yb$), one can notice that the same distribution considered in the previous paragraph (a non-local behavior for which Bob's box outputs $b=1$ \textit{almost always}) allows Bob to  control  the value of $f(y,b) = yb$. However, hidden signaling of $yb$ is not enough to reproduce non-local correlations in this scenario. To see this, notice that to satisfy the PR condition $a = xy \oplus b$ in all rounds, Alice's box must have access to both $y$ and $b$. In the previous cases, this was accomplished using the signaled bit and the shared bit $\lambda$, but given that the AND function does not always contain one bit of information (consider the case when $f(y,b) = yb = 0$), it is not possible for Alice's box to output $a= xy \oplus b$ in all rounds. Thus, it is not necessary to consider this case. 
However, it will be important to keep in mind that there are non-local quantum behaviors in which Bob has control over the value of the AND function of his local variables. 

\subsection{More general functions}
\label{sec:general}

We have considered all one bit functions, now we will show that our main result still holds in the case of a general function $f: \{ 0,1 \} ^2 \rightarrow \{ 0,1 \}^2$.

First, recall that when Hypothesis \ref{hyp:first} holds, for each $f$ we obtain a different distribution $p\left(a_n | x_n , x_{n-1}, f(y_n,b_n ), f(y_{n-1},b_{n-1} ) \right)$. 
Now, notice that the set of four possible values for Bob's variables, $\{ y,b \} \in \{ 0,1 \}^2$, splits into at least two subsets, one for each value of $f(y_{n-1},b_{n-1})$. 
Since the partition induced by $f$ is always a refinement of this partition, and both partitions consist of two subsets, they necessarily match in the one bit function case.

%If hidden signaling of such a function, along with a shared hidden variable, is enough to reproduce non-local correlations, then when hypothesis \ref{hyp:first} holds, we know that the set of four possible values for Bob's variables, $\{ y,b \} \in \{ 0,1 \}^2$, splits in at least two subsets, one for each value of $f(y_{n-1},b_{n-1})$ that gives a different $p\left(a_n | x_n , x_{n-1}, f(y_n,b_n ), f(y_{n-1},b_{n-1} ) \right)$. Note that the partition induced by $f$ is always a refinement of this partition, and they necessarily match in the one bit function case (because both partitions consist of two subsets). Also note that the functions we have already considered, correspond to all possible partitions of $\{ 0,1 \}^2$ into two subsets.

The remaining functions we need to consider correspond to partitions of $\{ 0,1 \}^2$ into more than two subsets, and hence they are refinements of the ones studied in the previous sections. Therefore, Alice can use memory effects to gather \emph{at least} the same  information of Bob's variables in the previous round. There is one extra issue in this case that is not present when the signal is a one-bit function. Since a general function can be thought of as the refinement of some one-bit function, without loss of generality we can assume that once we pick a given behavior, the memory hypothesis ensures that:
\begin{equation}\label{eq:sesgogeneral}
\begin{split}
p ( a_{2n+1} = 0 | \tilde{x} , \tilde{\tilde{x}} , \tilde{y} , \tilde{b}, g_{2n} = 0 ) & = \alpha \, , \\
p ( a_{2n+1} = 0 | \tilde{x} , \tilde{\tilde{x}} , \tilde{y}, \tilde{b}, g_{2n} = 1 ) & = \beta \, ,
\end{split}
\end{equation}
where $\alpha \neq \beta$; $g_{2n}$ is the value of some one-bit function on even rounds; $\tilde{x}$, $\tilde{\tilde{x}}$ are fixed values for Alice's inputs on two consecutive rounds; and $\tilde{y}$, $\tilde{b}$ are fixed values for Bob's variables on odd rounds. Notice that $\tilde{y}$, $\tilde{b}$ are the values required to fix $f(y,b)$ on odd rounds and they are not necessarily unique (in fact, this is the case only when $f$ contains the two bits of information from Bob's side), but we need to assume this to cover all  cases. Notice also that the one-bit function $g$ in the last expression depends upon the behavior considered. This is  due to the fact that the uniformity assumption ensures that $f$ is the same for all behaviors, however, when $f$ contains more than one bit of information there is more than one non-trivial coarse grained versions of it.

% As we said before, for two-set partitions of Bob's variables (one-bit functions of $y$ and $b$), we can always find a non-local and quantum behavior $p(ab|xy)$ that allows Bob to have control over which of the two partitions will contain the values of his local variables. Therefore, we can always find situations in which Bob can freely manipulate the value of $f(y_{n-1},b_{n-1}$) which has an effect on the statistics for $a_n$, but it is not necessarily true that he will be able to fix the value of $f(y_n,b_n)$ which appears on Hypothesis \ref{hyp:first} (this was possible in the one-bit case because the two partitions were the same). If this were the case, then the protocol presented in previous section could fail, but this problem can be avoided considering a new, sufficiently close, behavior for which Bob can force some bias on the value of $f(y_n,b_n)$. For example, if Hypothesis \ref{hyp:first} holds with $f(y_n,b_n) = b_n$ and we are considering a behavior $p(ab|xy)$ for which $p(b|y) =1/2 \, \forall y$, the protocol presented is useless for communication, given that Bob could not fix the value of $f$ on odd rounds;

There are two subtleties in the general case that we would like to discuss here. First, it may happen that the 
considered behavior  is such that Bob has no control over the value of $g_{2n}$ in Eqs.~\eqref{eq:sesgogeneral}. For instance, if Bob's output is the one bit function but the round to round distribution is such that $p(b|y) =1/2 \, \forall y$, the protocol we presented in the previous sections is useless for communication, given that Bob could not codify his message in the value of $g_{2n}$. If this were the case, we can consider a quantum and non local behavior of the form
\begin{equation}
(1-\mu) p(ab|xy) + \mu p_v (ab|xy) \, ,
\end{equation}
where $ p_v (ab|xy)$ is the same local vertex considered in Sec. \ref{sec:output}, then for any $\mu \in (0,1]$ Bob can induce some bias on the values of $g_{2n}$ on even rounds. Moreover, if $\mu$ is small enough, Hypothesis \ref{hyp:first} still ensures a finite difference between the two lines in Eq.~\eqref{eq:sesgogeneral} for this new behavior.
Then, following the same protocol as before, Alice's local statistics on a given round will have a non trivial dependence on Bob's inputs of the previous round, and a contradiction with special relativity immediately follows. Of course, an analogous argument can be given in the case of any one bit function $g$, because we know that for all one-bit functions, there are local vertices for which Bob has control over its values.

The other issue that could appear is the following: Bob can choose $\tilde{y}$ as input in odd rounds, but he may not be able to fix $\tilde{b}$ as  output in  these rounds. In most cases, one would expect that this is not a problem because they have sampled the probability distributions. Thus, despite not having control over the value of Bob's outputs, they know $p(\tilde{b} | \tilde{y})$, from which they can reduce Eqs.~\eqref{eq:sesgogeneral}:
\begin{equation}
\begin{split}
p ( a_{2n+1} = 0 | \tilde{x} , \tilde{\tilde{x}} , \tilde{y} ,  g_{2n} = 0 ) & = \tilde{\alpha} \, , \\
p ( a_{2n+1} = 0 | \tilde{x} , \tilde{\tilde{x}} , \tilde{y},  g_{2n} = 1 ) & = \tilde{\beta} \, .
\end{split}
\end{equation}
If these new probabilities $\tilde{\alpha}$ and $\tilde{\beta}$ are different, then they can proceed as before and use the devices for communication. If it were the case that the behavior  is such that the two marginals coincide ($\tilde{\alpha} = \tilde{\beta}$), we can use the same trick as in the previous example: consider another sufficiently close behavior, for which the bias on $p(\tilde{b} | \tilde{y})$ is large enough to take advantage of Eqs.~\eqref{eq:sesgogeneral} for communication. It is important to note that these two issues can  appear only if fine tuning \cite{wood2015lesson} takes place, so our result is robust in the sense that holds for almost any non-local and quantum behavior.

This finally shows that hidden signaling of an arbitrary function of Bob's variables plus Hypothesis \ref{hyp:first}, leads to a conflict with causality.

%%% SAMPLING PROTOCOL
\section{Sampling and signalization protocol}\label{sec:protocolo}

In the previous sections we showed that, as long as Alice and Bob are able to infer the probability distribution $p\left(a_{n} | x_{n}, x_{n-1} ,  f(y_n,b_n ), f(y_{n-1},b_{n-1} ) \right)$, Hypothesis~\ref{hyp:first} cannot hold without contradicting special relativity. To complete our argument, we must show that it is in fact possible for Alice to learn how her outputs are biased by Bob's variables.

First, it is important to note that, \textit{a priori}, Alice and Bob do not know which function $f(y,b)$ is being hiddenly signaled between their boxes. As we will see, in general they cannot directly access to this information, however they can devise a strategy to use it for communication. The main idea behind the sampling protocol is quite simple: if Alice and Bob choose their inputs randomly, and Bob sends his local information through a classical channel to Alice, then she can sample $p (a_{n} | x_{n},y_n,b_n , x_{n-1},  y_{n-1},  b_{n-1})$. Note that there are $64$ of these probabilities which are relevant (but it is enough to sample them with $a_n=0$ , since the remaining ones can be obtained by normalization). It is useful to set the notation $G_{y_{n-1} , b_{n-1}} ( x_n , x_{n-1} , y_n , b_n ) \equiv p (a_{n} = 0 | x_{n},y_n,b_n , x_{n-1},  y_{n-1},  b_{n-1}) $. Hypothesis \ref{hyp:first} then indicates that there is at least one choice of $y,b,\tilde{y},\tilde{b}$ such that $f(y,b) \neq f(\tilde{y},\tilde{b})$ and
\begin{equation}
G_{y , b } \neq G_{ \tilde{y} , \tilde{b} } ,
\label{hypG}
\end{equation}
as functions of their four arguments. Now we see why, in general, Alice and Bob cannot decide which function $f(y,b)$ is being signaled between the boxes. Equation \eqref{hypG} does not necessarily hold \emph{for all} choices of $y,b,\tilde{y},\tilde{b}$ when  $f(y,b) \neq f(\tilde{y},\tilde{b})$, so the sampling protocol only allows them to obtain a \textit{coarse-grained} version of $f(y,b)$ (just the instances of it carrying a detectable memory effect on Alice's side). In other words, the sampling protocol allows them to distinguish the partitions of the set $\{0,1\}^2$ induced by the memory effects but not necessarily the finer partitions induced by the signaled function. 

Now, if there were no memory, we should have
\begin{equation}
G_{y_{n-1} , b_{n-1}} ( x_n , x_{n-1} , y_n , b_n ) = p(a_n = 0 | x_n y_n b_n) \, 
\label{Gsinmemoria}
\end{equation}
for all $y_{n-1}$, $b_{n-1}$ and $x_{n-1}$. The right hand side of the last equation can  easily be obtained from the round to round distribution and is, therefore, known by Alice. The protocol is then the following: First, there is a learning stage in which Alice and Bob pick their inputs randomly and share their results through a classical channel. As rounds go by, given that their inputs are being chosen randomly, all possible sequences of outcomes will occur and they can estimate each $G_{y_{n-1} , b_{n-1}} ( x_n , x_{n-1} , y_n , b_n )$ as the relative frequencies of each string. The difference between their estimation and the actual value of $G_{y_{n-1} , b_{n-1}} ( x_n , x_{n-1} , y_n , b_n )$ can be easily bounded and is a decreasing function of the number of rounds they have shared. When the number of shared rounds is such that the error is small enough to be sure that a discrepancy was found between the actual functions $G_{y_{n-1},b_{n-1}}$ and their expected values in the memoryless case (see Eq.~\eqref{Gsinmemoria}), then they are in a position to start applying the signalization protocol.
This protocol was mentioned briefly in the previous sections, and we will proceed to describe it in more detail.

As we have discussed in Sec.~\ref{sec:general}, the specific function that is being signaled is not relevant to this analysis, if Hypothesis \ref{hyp:first} holds, there is always a non-local and quantum behavior $p(ab|xy)$ for which
\begin{equation}
\begin{split}
p ( a_{2n+1} = 0 | \tilde{x} , \tilde{\tilde{x}} , \tilde{y} , y_{2n} = 0 ) & = \alpha \, , \\
p ( a_{2n+1} = 0 | \tilde{x} , \tilde{\tilde{x}} , \tilde{y} , y_{2n} = 1 ) & = \beta \, ,
\end{split}
\end{equation}
where $\tilde{x}$, $\tilde{\tilde{x}}$ and $\tilde{y}$ are fixed input choices for $x_{2n+1}$, $x_{2n}$ and $y_{2n+1}$ respectively, and $\alpha \neq \beta$ are two probabilities known by Alice and Bob after the learning stage. 
If Bob chooses $\tilde{y}$ as his input in odd rounds and always his secret message as his input in the even ones, and Alice sticks to her task of choosing $\tilde{x}$ and $\tilde{\tilde{x}}$ as her inputs in odd and even rounds respectively, then she should expect to see outputs with a fraction $\alpha$ of zeros in the odd rounds if Bob's message $y=0$, and a fraction $\beta$ of zeros when Bob's message is $y=1$.
%If Alice stick to her task of choosing $\tilde{x}$ and $\tilde{\tilde{x}}$ as her inputs in odd and even rounds respectively, and Bob chooses $\tilde{y}$ as his input in odd rounds and always his secret message as his input in the even ones, Alice should expect to see outputs with a fraction $\alpha$ of zeros in the odd rounds if Bob's message $y=0$, and a fraction $\beta$ of zeros when Bob's message is $y=1$.
At this point, it becomes clear that this procedure allows her to guess something about Bob's secret message with some probability of success different from $1/2$ (which already triggers a contradiction with causality). But now we will see that, in fact, they can devise a strategy to communicate with probability of success arbitrarily close to $1$. 

\begin{figure}
    \centering
    \includegraphics[width=1\columnwidth]{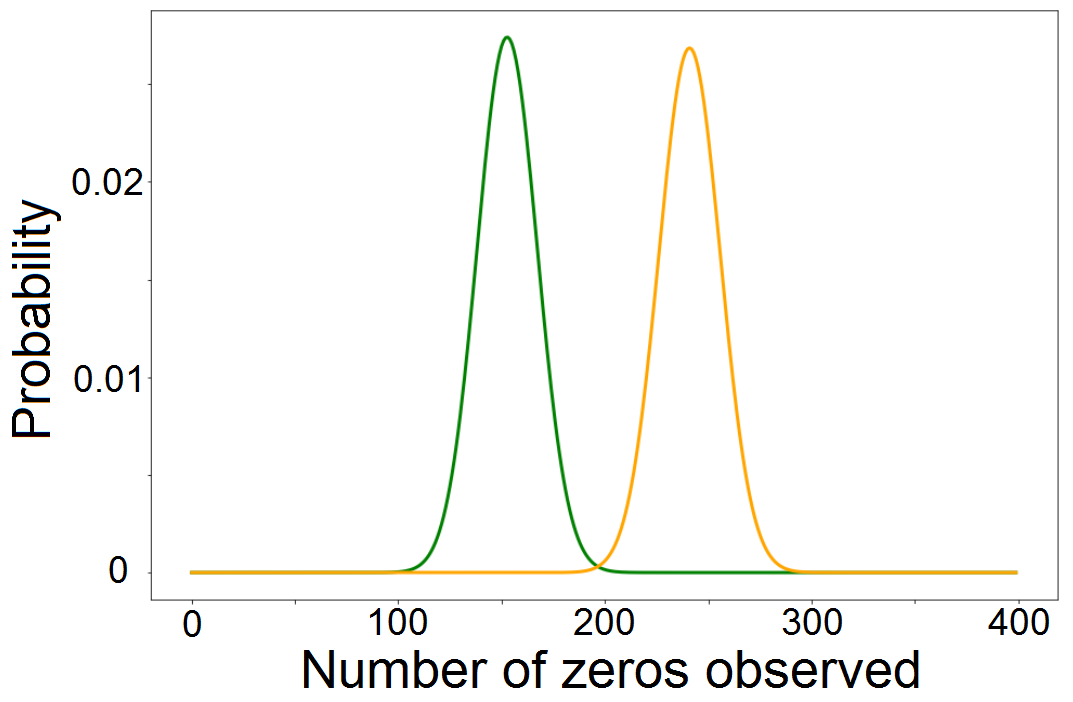}
    \caption{Binomial distributions with parameters $p=\alpha=0.4$ and $N=882$ (green line); and $p=\beta=0.5$ and also $N=882$ (orange line). This value of $N$ is obtained from Eq.~\eqref{Nprotocolo} for $k=3$. If, out of her $N=882$ odd round outcomes she observes a number of zeros below the intersection point at the center, she has a $99.7 \%$ confidence level that Bob is not entering $y=1$ as his input on even rounds. Analogously, she can discard $y=0$ as Bob's input with the same confidence level when the number of zeros observed is above that limit. } 
    \label{fig:binomiales}
\end{figure}

Let us consider that Alice and Bob shared $2N$ rounds under the signalization protocol, then Alice's outputs in odd rounds will be equivalent to $N$ tosses of a biased coin. 
Except that if Bob is trying to send her the bit $y=0$  the fraction of zeros for this coin would be $\alpha$, and will be $\beta$ for $y=1$. Intuitively, it is clear that Alice's certainty about which coin is being tossed 
will increase with the number of rounds and also with the difference between $\alpha$ and $\beta$.
Notice also that the distribution of Alice's outputs in odd rounds is given by binomial distributions with success probabilities $\alpha$ and $\beta$, respectively, both with a number of trials equal to $N$. For a binomial distribution with parameters $N$ and $p$, the mean value and standard deviation are $Np$ and $\sqrt{Np(1-p)}$ respectively.  If, without loss of generality, we assume that $\alpha < \beta$, we can fix $N$ as the smallest natural number such that
\begin{equation}
N \alpha + k \sqrt{N\alpha (1-\alpha)} < N \beta - k \sqrt{N\beta (1-\beta)} \, ,
\end{equation}
where $k$ is a positive number that quantifies the overlap between both distributions. By equating the expression, it is easy to see that $N$ should be the smallest natural number such that
\begin{equation}
N > \frac{k^2 \left( \sqrt{ \alpha (1-\alpha) } + \sqrt{\beta (1-\beta)} \right)^2 }{(\beta - \alpha)^2} \, . 
\label{Nprotocolo}
\end{equation}
As $k$ increases, so do $N$ and Alice's ability to distinguish between coins. For instance, for $k=3$ we have the higher tail and lower tail of each distribution intersecting at three standard deviations from each mean value. An example is plotted in Fig.~\ref{fig:binomiales}. For large $N$, this distribution can be approximated by a normal distribution with the same mean value and standard deviation. So, if we fix $N$ using the previous equation for $k=3$, when the fraction of zeros observed by Alice lies in the $\alpha$ ($\beta$) region, she has a confidence level of $99.7 \% $ that Bob is not pressing $y=0$ ($y=1$) in his even rounds. Of course, this confidence level can be made as large as they want by increasing the value of $N$. If they follow the signalization protocol, then Alice can infer Bob's input in even rounds, achieving one bit of communication after the $2N$ rounds. 

Finally, to complete our argument,  we can mention that the conflict with causality comes from the fact that the distance between Alice and Bob, $d$, and the interval between rounds, $\tau$, can always be fixed in such a way that
$\frac{d}{c} > 2 N  \tau$, where $c$ is the speed of light. Meaning that the speed of the information Bob is sending to Alice is a faster-than-light.

We have shown that deterministic models of non-local correlations cannot have memory, in the sense of Hypothesis \ref{hyp:first}, otherwise it will lead to a contradiction with special relativity. Our argument can be easily generalized to the case of memory of more than one round as follows. If Alice's box has memory of $k$ rounds, we have that $p(a_n | x_n ,\dots ,x_{n-k} , f(y_n,b_n) , \dots , f(y_{n-k} , b_{n-k} ) )$ has a non-trivial dependence on $f(y_{n-1},b_{n-1}),  \dots , f(y_{n-k} , b_{n-k} ) $.  Once the sampling protocol is completed and Alice and Bob are aware of this bias, the signalization protocol is  basically the same as before: They agree to fix the values of $x_{n} , \dots , x_{n-k}$ and the value of $f(y_{n},b_{n})$; and then Bob can encode the message in the values of $f(y_{n-1},b_{n-1}),  \dots , f(y_{n-k} , b_{n-k} )$. 

%%% SUMMARY
\section{Summary}

In this paper, we have studied the implications of memory effects in hidden signaling models for non-local correlations in the $(2,2,2)$ scenario. We showed that if it were the case that nature resorts to hidden signaling with the aid of memory to generate these kinds of correlations, the agents would be able to achieve faster-than-light communication. Of course, this result does not rule out hidden signaling and, strictly speaking, we showed that if it existed, then memory effects would have to be forbidden by nature in these kinds of scenarios. As memory is a resource easily available in nature, it seems reasonable to think of this result as theoretical evidence against hidden signaling as an explanation of non-local quantum correlations.

%%%%%%% ACKNOWLEDGMENT
\section*{Acknowledgement}

This work was supported by ANPCyT (Grants PICT No.~2014-3711, No.~2015-2293, No.~2016-2697, and No.~2018-04250), CONICET-PIP No.~11220130100329CO, and UBACyT.
%%%%%%% BIBLIOGRAPHY
\bibliography{hiddensignalingnuevoenfoque}

%%%
\end{document}